\documentclass[preprints,article,accept,pdftex,moreauthors]{Definitions/mdpi} 

\usepackage{siunitx}
\usepackage{xcolor}
\usepackage{ulem}

\usepackage{graphicx}
\usepackage{subcaption}

\firstpage{1} 
\pubvolume{1}
\issuenum{1}
\articlenumber{0}
\pubyear{2023}
\copyrightyear{2023}
\datereceived{} 
\dateaccepted{} 
\datepublished{} 
\hreflink{https://doi.org/} 

\graphicspath{{fig/}}
\Title{Astrocyte control bursting mode of spiking neuron network with memristor-implemented plasticity}

\TitleCitation{Astrocyte control bursting mode of spiking neuron network with memristor-implemented plasticity}

\Author{Sergey V. Stasenko $^{1,2,4}$\orcidA{}, Alexey N. Mikhaylov $^{3,4}$\orcidB{}, Alexander A. Fedotov $^{4}$\orcidC{}, Vladimir A. Smirnov $^{4}$\orcidD{} and Victor B. Kazantsev $^{1,2,4,}$*\orcidE{}}

\AuthorNames{Sergey V. Stasenko, Alexey N. Mikhaylov, Alexander A. Fedotov, Vladimir A. Smirnov and Victor B. Kazantsev}

\AuthorCitation{Stasenko, S.V.; Mikhaylov, A.N.; Fedotov, A.A. ; Smirnov, V.A.; Kazantsev V.B.}

\address{%
$^{1}$ \quad Laboratory of neurobiomorphic technologies, Moscow Institute of Physics and Technology, 117303 Moscow, Russia\\
$^{2}$ \quad Laboratory of Advanced Methods for High-Dimensional Data Analysis, Lobachevsky State University of Nizhny Novgorod, 603022 Nizhny Novgorod, Russia; stasenko@neuro.nnov.ru (S.V.S.)\\
$^{3}$ \quad Laboratory of memristor nanoelectronics, Lobachevsky State University of Nizhny Novgorod, 603022 Nizhny Novgorod, Russia (mian@nifti.unn.ru)\\
$^{4}$ \quad Research Laboratory Neuroelectronics and Memristive Nanomaterials (NEUROMENA Lab), Institute~of~Nanotechnologies, Electronics and Electronic Equipment Engineering, Southern Federal University, 347922 Taganrog, Russia
}

\corres{Correspondence: kazantsev@neuro.nnov.ru}

\abstract{A mathematical model of a spiking neuron network accompanied by astrocytes is considered. The network is composed of excitatory and inhibitory neurons with synaptic connections supplied by a memristor-based model of plasticity. Another mechanism for changing the synaptic connections involves astrocytic regulations using the concept of tripartite synapses. In the absence of memristor-based plasticity, the connections between these neurons drive the network dynamics into a burst mode, as observed in many experimental neurobiological studies when investigating living networks in neuronal cultures. The memristive plasticity implementing synaptic plasticity in inhibitory synapses results in a shift in network dynamics towards an asynchronous mode. Next,it is found that accounting for astrocytic regulation in glutamatergic excitatory synapses enable the restoration of 'normal' burst dynamics. The conditions and parameters of such astrocytic regulation's impact on burst dynamics established.}



\begin{document}

\section{Introduction}
\label{sec:intro}

Bursting is a common neural phenomenon marked by synchronized spiking in neuron groups, enhancing synaptic reliability and selective communication within spiking neuron networks (SNNs) \cite{izhikevich2006bursting}. Controlling bursts is vital for accurate information processing and network stability \cite{izhikevich2003bursts}. The intricate patterns and temporal relationships in burst dynamics, as discussed by Sjöström et al. (2006) \cite{sjostrom2006cooperative}, provide insight into the properties of neurons and the interactions within networks. These dynamics are influenced by various factors, including excitatory and inhibitory connections, network topology, and individual neuron properties. The temporal aspects of bursting play a crucial role in time-sensitive learning and plasticity, facilitating the wiring of neurons based on their firing patterns, as explored by Buzsáki (1984) \cite{buzsaki1984feed} and Abeles (1991) \cite{abeles1991corticonics}.

In dissociated neuronal cultures, population bursts exhibit quasisynchronous dynamics, representing high-frequency spike sequences involving most network neurons within a defined time window. These bursts have distinct activation profiles, encoding various network dynamical states \cite{wagenaar2006extremely, wagenaar2005controlling, wagenaar2006persistent,pasquale2008self,zeldenrust2018neural, pimashkin2011spiking}.

Neuronal synchronization mechanisms can be categorized into some groups: neuron properties \cite{wang2010neurophysiological}, network properties  \cite{zeitler2009asymmetry,pikovsky2002synchronization,tsybina2022synchronization,simonov2014synchronization}, and neuromodulation \cite{muthukumaraswamy2009resting}  mediated by various neuromodulators. Most of these groups operate on a millisecond timescale, except for neuromodulation involving glial cells (astrocytes) \cite{stasenko2020quasi,
lazarevich2017synaptic, makovkin2022controlling,stasenko2023information}, which acts on a second timescale, and extracellular matrix molecular of the brain, which acts on a minutes and hours timescale \cite{kazantsev2012homeostatic,lazarevich2020activity,rozhnova2021bifurcation,stasenko2023bursting}.

In numerous neurobiological investigations, astrocytes play a crucial role in the regulation of synaptic information transmission \cite{Araque1998, Araque1999, Wittenberg2002, Wang1999}. Through the calcium-dependent release of neuroactive chemicals, termed gliotransmitters (such as glutamate, adenosine triphosphate (ATP), D-serine, GABA), astrocytes impact both the pre- and postsynaptic components of the synapse. These observations have given rise to the tripartite synapse concept \cite{Wittenberg2002, Araque1999, Haydon2001}, wherein astrocytes modulate synaptic transmission by interacting with presynaptic and postsynaptic receptors \cite{Perea2009, Martin2007}.

Mathematical models have been proposed to comprehend the functional role of astrocytes in neuronal dynamics, including concepts such as the "dressed neuron" \cite{Nadkarni2004, Nadkarni2007}, astrocytes acting as frequency-selective "gatekeepers" \cite{Volman2007}, and their influence on presynaptic functions \cite{DePitta2011}. Other studies have highlighted astrocytes' participation in spike-timing-dependent plasticity (STDP), learning \cite{Postnov2007, Amiri2011, Wade2011}, and neural synchrony \cite{amiri2013astrocyte, pankratova2019neuronal}. Moreover, mean-field models have been employed to phenomenologically describe astrocytic modulation of neuronal synchronization \cite{Gordleeva2012, de2019gliotransmitter, lenk2020computational, barabash2021stsp, stasenko20223d, barabash2023rhy, olenin2023dynamics}.

The interaction between astrocytic and synaptic elements exhibits variations depending on structural and functional coupling. In network dynamics, models of spiking neural networks (SNNs) accompanied by astrocytes have been developed \cite{oschmann2018silico, postnov2009dynamical, de2022multiple, stasenko2022astrocytes, stasenko2023dynamic}, demonstrating that astrocytes enhance short-term memory \cite{gordleeva2021modeling, zimin2023artificial}. Recent experiments suggest the involvement of astrocytes in up-down synchronization in neuronal networks, though the underlying mechanisms remain uncertain. Neural network models, incorporating excitatory neurons, inhibitory neurons, and astrocytes, have been created to investigate how astrocytes influence this phenomenon, revealing that astrocytes can promote realistic up-down regimes in neural networks \cite{blum2022modelling}.

Spike-timing-dependent plasticity (STDP) \cite{vogels2011inhibitory, wu2022regulation}, is another factor that contributes to neural synchronization. It involves the adjustment of excitatory and inhibitory connection strengths \cite{froemke2007synaptic}, helping the network adapt to input patterns and learning demands.

Neuronal synchronization is critical for cognitive functions like attention, perception, and decision-making \cite{haider2006neocortical, atallah2012parvalbumin, maimon2006cognitive,stasenko2023loss}. Understanding their mechanisms also sheds light on neurodevelopmental and neuropsychiatric disorders such as autism and schizophrenia \cite{howes2022integrating, sohal2019excitation}, offering potential therapeutic insights.

Memristive plasticity, as studied in \cite{yang2013memristive,kipelkin2023mathematical}, holds promise in information processing and computing. Memristors can change their resistance in response to electrical signals, akin to biological synaptic plasticity \cite{pershin2011memory}. These memristive devices are being explored for various computing purposes, including memory, logic gates, and neuromorphic synapse-like components \cite{jo2010nanoscale}. This approach offers the potential for efficient and scalable computing platforms mimicking biological neural networks \cite{indiveri2011neuromorphic}. Memristive plasticity relies on the conductance changes in memristors, driven by mechanisms like charged defect migration or material restructuring. These changes enable adaptive synaptic weight adjustments, resembling long-term potentiation (LTP) and long-term depression (LTD) processes seen in neural networks \cite{pershin2011memory}. The possibilities of memristive plasticity in computational systems are vast, offering parallelism, non-volatility, and energy efficiency \cite{yang2013memristive}. Incorporating memristors into neural network models seeks to mimic synaptic plasticity and investigate hardware implementations for neuromorphic computing \cite{indiveri2011neuromorphic}.

This paper introduces a novel model for controlling burst dynamics in spiking neural networks with memristor-mediated plasticity. In the absence of this plasticity, the network exhibits bursting behavior. However, by incorporating memristive plasticity in inhibitory connections, the network transitions to asynchronous spiking. Investigating astrocytic regulation of glutamatergic synapses restores burst dynamics, highlighting the potential to switch the network's dynamics between modes.

The structure of this work is as follows: The Introduction section (\ref{sec:intro}) provides a concise overview of the research's relevance and the current state of the field. In the Model section (\ref{sec:model}), we delve into details about the neuron model, astrocyte dynamics, neuron-glial interactions, memristive plasticity, neural network structure, and the methodologies and libraries utilized to obtain our results. The Results section (\ref{sec:results}) highlights the study's main findings, showcasing the observed effects. The Discussions (\ref{sec:discussion}) and Conclusions (\ref{sec:conclusion}) sections outline potential research directions and offer a brief summary of the research's key outcomes.

\section{The model}
\label{sec:model}

\subsection{Neuron model}
To simulate the dynamics of a neuron, the integrate-and-fire neuron model was used \cite{dayan2005theoretical}. The integrate-and-fire neuron model is a simplified mathematical description of a neuron's behavior in the brain. It is a classic model used to study the basic principles of neuronal firing and information processing. The model simplifies the complex biophysical processes that occur in real neurons and focuses on the concept of integrating incoming electrical signals and generating an output spike when a certain threshold is reached. The integrate-and-fire model does not capture the detailed dynamics of action potentials, like the Hodgkin-Huxley model, but it simplifies the process to a binary "fire" or "not fire" event when the threshold is crossed. This model is useful for understanding the principles of neural information processing, such as how the integration of synaptic inputs leads to the generation of action potentials. It is also computationally efficient and can be used in large-scale neural network simulations.

In order to simplify the dynamics of synaptic connections between neurons, excitatory ($g_{exc}$) and inhibitory ($g_{inh}$) conduction-based synapses were used in the neuron model. The complete system of equations can be written as follows:

\begin{equation}
\begin{cases}
C_m \dfrac{dV}{dt}= - g_l(V - E_l) - (g_{exc} V + g_{inh} (V-E_r)) + I_{ext}, \\\\
\dfrac{dg_{exc}}{dt}=-\dfrac{g_{exc}} {\tau_{exc}} +\sum_i w^{i}_{exc} \cdot \delta\left(t-t_{\mathrm{spike}}^i \right),
\\\\
\dfrac{dg_{inh}}{dt}=-\dfrac{g_{inh}} {\tau_{inh}} +\sum_i w^{i}_{inh} \cdot \delta\left(t-t_{\mathrm{spike}}^i \right).
\end{cases}
\label{eq:neuron}
\end{equation}

Here, $V$ represents the membrane potential, $E_l$ denotes the leakage potential, $E_r$ is the reverse potential, $I_{ext}$ signifies the direct input current, $C_m$ stands for membrane capacity, $\tau_{exc}$ and $\tau_{inh}$ represent the time constants for excitatory and inhibitory synaptic inputs, respectively. Additionally, $g_{exc}$ and $g_{inh}$ are the synaptic conductances for excitatory and inhibitory synapses, and $w^{i}{exc}$ and $w^{i}{inh}$ are the weights of the excitatory and inhibitory synapses originating from neuron $i$. The variable $t_{\mathrm{spike}}^i$ denotes the time of spike occurrence in neuron $i$. The right side of the equation includes a summation term that accounts for synaptic activations resulting from presynaptic spikes. In the model described by Equation \ref{eq:neuron}, when the membrane potential reaches the threshold value $V_t$, the membrane potential resets to $E_l$.

In our model, synaptic weights between inhibitory neurons are fixed, between excitatory and inhibitory neurons and excitatory and excitatory neurons are regulated by astrocytes, and between inhibitory and excitatory neurons, synaptic weights are plastic and determined by memristor implemented synaptic plasticity.

\subsection{Memristor implemented synaptic plasticity}

The model of plasticity for nanocomposite memristors, denoted as $(\mathrm{CoFeB}){x}\left(\mathrm{LiNbO}{3}\right)_{1-x}$ \cite{demin2021necessary}, was utilized to represent synaptic plasticity in connections between inhibitory and excitatory neurons, as outlined:

\begin{equation}
\Delta w(\Delta t)= \begin{cases}A^{+} \cdot w \cdot\left[1+\tanh \left(-\frac{\Delta t-\mu_{+}}{\tau_{+}}\right)\right] \quad \text { if } \Delta t>0 ; \\ A^{-} \cdot w \cdot\left[1+\tanh \left(\frac{\Delta t-\mu_{-}}{\tau_{-}}\right)\right] \quad \text { if } \Delta t<0 .\end{cases}
\label{eq:memristive_plast}
\end{equation}

The constants remain unchanged, consistent with the results reported in the initial investigation \cite{demin2021necessary}: $A^{+}=0.074, A^{-}=-0.047$, $\mu^{+}=26.7 \mathrm{~ms}, \mu^{-}=-22.3 \mathrm{~ms}, \tau^{+}=9.3 \mathrm{~ms}, \tau^{-}=10.8 \mathrm{~ms}$.

Therefore, the alteration in the synaptic weight of the inhibitory synapse connecting inhibitory and excitatory neurons during the generation of a spike on a presynaptic neuron can be described as $w_{inh} \leftarrow w_{inh}+ \Delta w_{inh} $.

\subsection{Neuron-glial interaction}

In our neuron model, the occurrence of each spike leads to the release of a neurotransmitter. We employ glutamatergic synapses to illustrate neuron-glial interactions. Based on prior experimental and modeling investigations, we postulated that the mechanism of glutamate-mediated metabolism is the pivotal factor responsible for the synchronized firing of neurons \cite{Angulo2004, Halassa2009}.

In order to simplify, we will adopt a phenomenological model to characterize the dynamics of released glutamate. Utilizing a mean-field approximation, we derived the mean extrasynaptic concentration of glutamate for each excitatory synapse, identified as $X$, through the following equations:

\begin{equation}
	\dfrac{dX}{dt} = -\dfrac{X}{\tau_X} + b_X \theta\left(t-t_{\mathrm{spike}}^i \right).
	\label{eq:glutamate}
\end{equation}

Here, $b_X$ signifies the glutamate release fraction, and $\tau_{X}$ denotes the relaxation time. When a presynaptic neuron produces a spike, neurotransmitter is released, leading to a rise in synaptic neurotransmitter concentration. This concentration subsequently diminishes over time, as described by $\tau_{X}$. The parameter values employed are $\tau_{X} = 20 \mathrm{~ms}$ and $b_X=1$.

A portion of the synaptic glutamate can bind to the metabotropic glutamate receptors on astrocyte membrane. Following a cascade of molecular changes initiated by an increase in intracellular calcium, the astrocyte releases gliotransmitter into the extracellular space. In our mathematical model, we simplified the representation by excluding detailed descriptions of these transformations. Instead, we established an input-output functional relationship between neurotransmitter and gliotransmitter concentrations, expressed as \cite{stasenko2020quasi, Gordleeva2012, lazarevich2017synaptic}:

\begin{equation}
	\dfrac{dY}{dt} = - \alpha_{Y}Y + \dfrac{\beta_{Y}}{1+exp(-X + X_{thr})}. \label{eq:astrocyte}
\end{equation}

In the equation \label{eq:astrocyte}, $Y$ denotes the concentration of gliotransmitter near the corresponding excitatory synapse, and $\alpha_{Y}$ stands for the clearance rate. The parameter values employed are $\alpha_{Y} = 120 \mathrm{~ms}$, $\beta_{Y} = 1$, $X_{thr} = 4$.

The second component in Equation (\ref{eq:astrocyte}) accommodates the production of gliotransmitter when the mean-field gliotransmitter concentration exceeds the specified threshold, $X_{thr}$.

Empirical observations indicate that astrocytes have the potential to influence neurotransmitter release probability, leading to the possible induction of synaptic potentiation or depression and, consequently, modulation of synaptic currents \cite{Perea2007, Jourdain2007, Fiacco2004}. In our model, we concentrate on the enhancement of synaptic transmission (an increase in neurotransmitter release probability) and incorporate it in the following manner for glutamatergic synapses during the generation of a spike in a presynaptic neuron:

\begin{equation}
	w_{exc} \leftarrow w_{exc}+ \Delta w_{exc}(1 + \dfrac{\gamma_{Y}}{1 + exp(-Y + Y_{thr})} ). \label{eq:astrocyte_influence}
\end{equation}

Here, $w_{exc}$ represents the weight for glutamatergic synapses between neurons, and $\gamma_{Y}$ is the coefficient reflecting the influence of astrocytes on synaptic connections.

\subsection{Neural Network}

\begin{figure}[H]
   \centering
  \includegraphics[width=0.95\textwidth]{
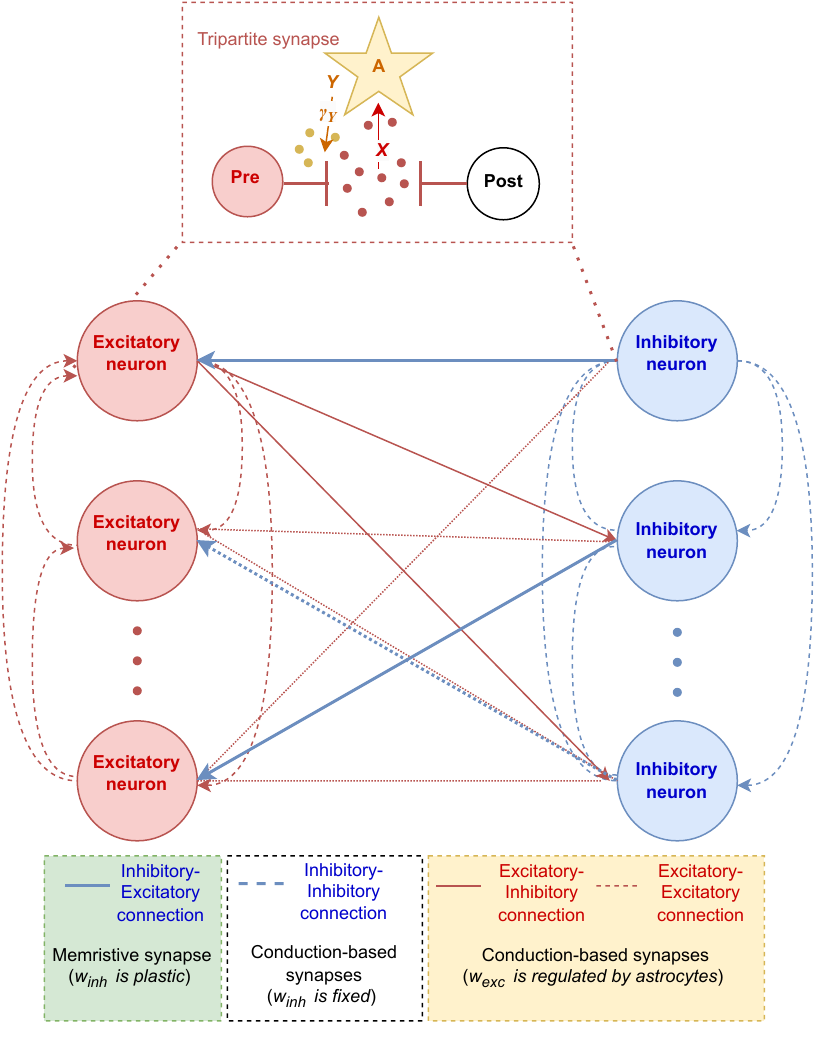}
  \caption{Scheme of spiking neuron network with memristive plasticity and astrocyte regulation of synaptic connection. Symbols on the scheme: $A$ - astrocyte, $Pre$ and $Post$ - pre and postsynaptic neurons, $X$ - neurotransmitter (glutamate), $Y$ - gliotransmitter (glutamate), $\gamma_Y$ - the coefficient of astrocyte influence on synaptic connection.}
  \label{fig:SNN_scheme}
\end{figure}

The diagram in Figure \ref{fig:SNN_scheme} illustrates the configuration of the spiking neuron network. This network comprises 8000 excitatory neurons (depicted in red) and 2000 inhibitory neurons (depicted in blue), interconnected in an "all-to-all" manner with a connection probability of 2 percent. In this network, inhibitory-excitatory connections, represented by bold blue arrows, are established using synapses with memristive plasticity. The figure also includes a schematic representation of the neuron-glial interaction implemented for glutamatergic synapses. The remaining connections, including excitatory-excitatory, excitatory-inhibitory, and inhibitory-inhibitory connections, are formed using synapses with dynamics governed by Equations \ref{eq:neuron} and \ref{eq:astrocyte_influence} (for excitatory-excitatory and excitatory-inhibitory connections).

\subsection{Numerical simulation and data analysis methods}

We performed numerical calculations using the Euler method with a 0.01 integration step. Our Python-based computational program, utilizing Pandas \cite{nelli2015pandas} for data processing and analysis, was employed for the model. Simulations were executed with Brian2 \cite{stimberg2019brian}, and data visualization and analysis were facilitated by the Matplotlib, Seaborn \cite{bisong2019matplotlib}, and Scipy \cite{virtanen2020scipy} libraries.

Population activity rates were determined by summing neuron spikes per second, and smoothing was achieved using a 0.5 ms standard deviation Gaussian window.

\section{Results}
\label{sec:results}

To study the astrocytic regulation of the spiking neural network with memristor-based plasticity on the processes of neuronal synchronization, we considered several cases:
\begin{itemize}
 \item Synchronization of neurons in the spiking neuron network without any regulation;
 \item Suppression of neuronal synchronization in the presence of memristor-based plasticity;
 \item Restoration of neuronal synchronization in the presence of memristor-based plasticity and astrocytic regulation of synaptic transmission.
\end{itemize}

To explore the final two cases, we employed the following simulation protocol for the model:
\begin{itemize}
 \item In the initial second, the spiking neuron network model was computed without memristor-based plasticity;
 \item Following the first second, memristor-based plasticity was engaged in the synapses connecting inhibitory and excitatory neurons.
\end{itemize}

\begin{figure}[H]
	\centering
	\textit{(a)}\includegraphics[width=0.95\textwidth]{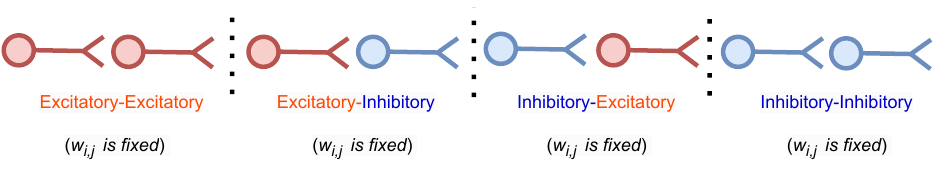}
	\textit{(b)}\includegraphics[width=0.95\textwidth]{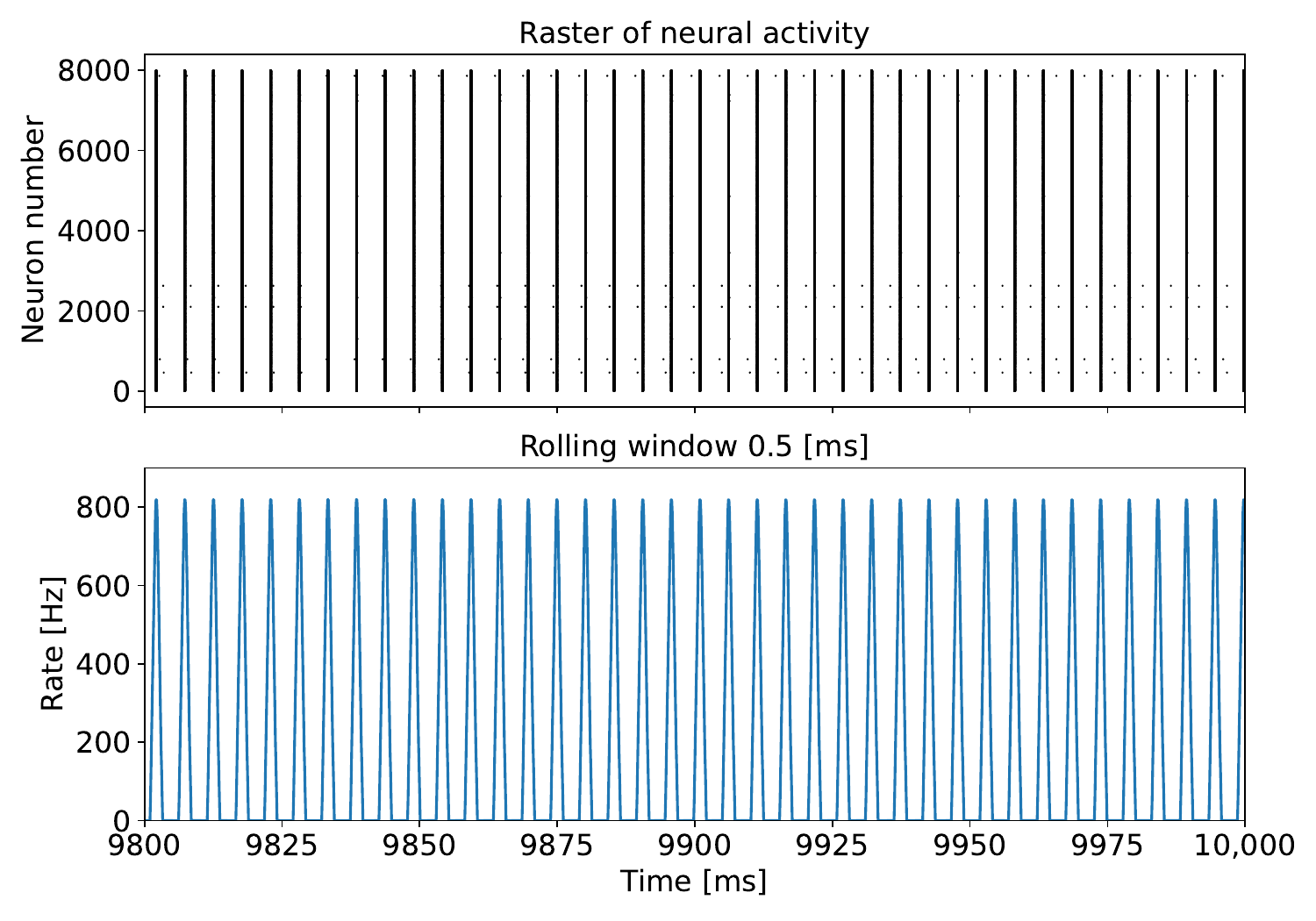}
	\caption{ a) Diagram illustrating the synaptic contacts of spiking neuron network in the case of fixed synaptic weights. Excitatory neurons are highlighted in red, while inhibitory neurons are highlighted in blue. $w_{i,j}$- synaptic weight. b) Raster diagram depicting neuronal activity and the population activity rate in the scenario where neurons in the spiking neuron network synchronize without any regulation.}
	\label{fig:raster_without_all}
\end{figure}

The total duration of the model simulation was 10 seconds. The first case involved examining the dynamics of a spiking neuron network in the absence of memristive plasticity and astrocytic regulation. To accomplish this, consider a neural network with fixed synaptic weights (see Figure \ref{fig:raster_without_all}a). The Figure \ref{fig:raster_without_all}b illustrates the network's dynamics during the simulation period from 9.8 to 10 seconds. This specific visualization interval was selected to facilitate the observation of network dynamics and to minimize the influence of any potential transient processes at the beginning of the simulation. As shown, neurons exhibit synchronization with the formation of regular burst activity, a phenomenon observed in numerous experimental neurobiological studies exploring living networks in neuronal cultures \cite{pimashkin2011spiking}.

\begin{figure}[H]
	\centering
	\textit{(a)}\includegraphics[width=0.95\textwidth]{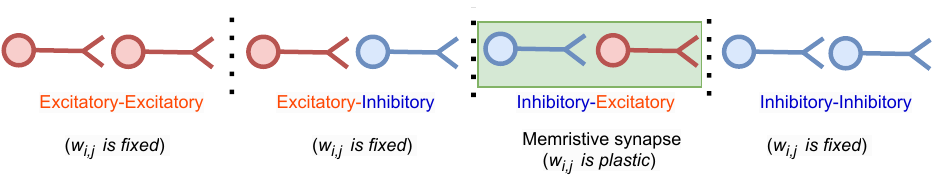}
	\textit{(b)}\includegraphics[width=0.95\textwidth]{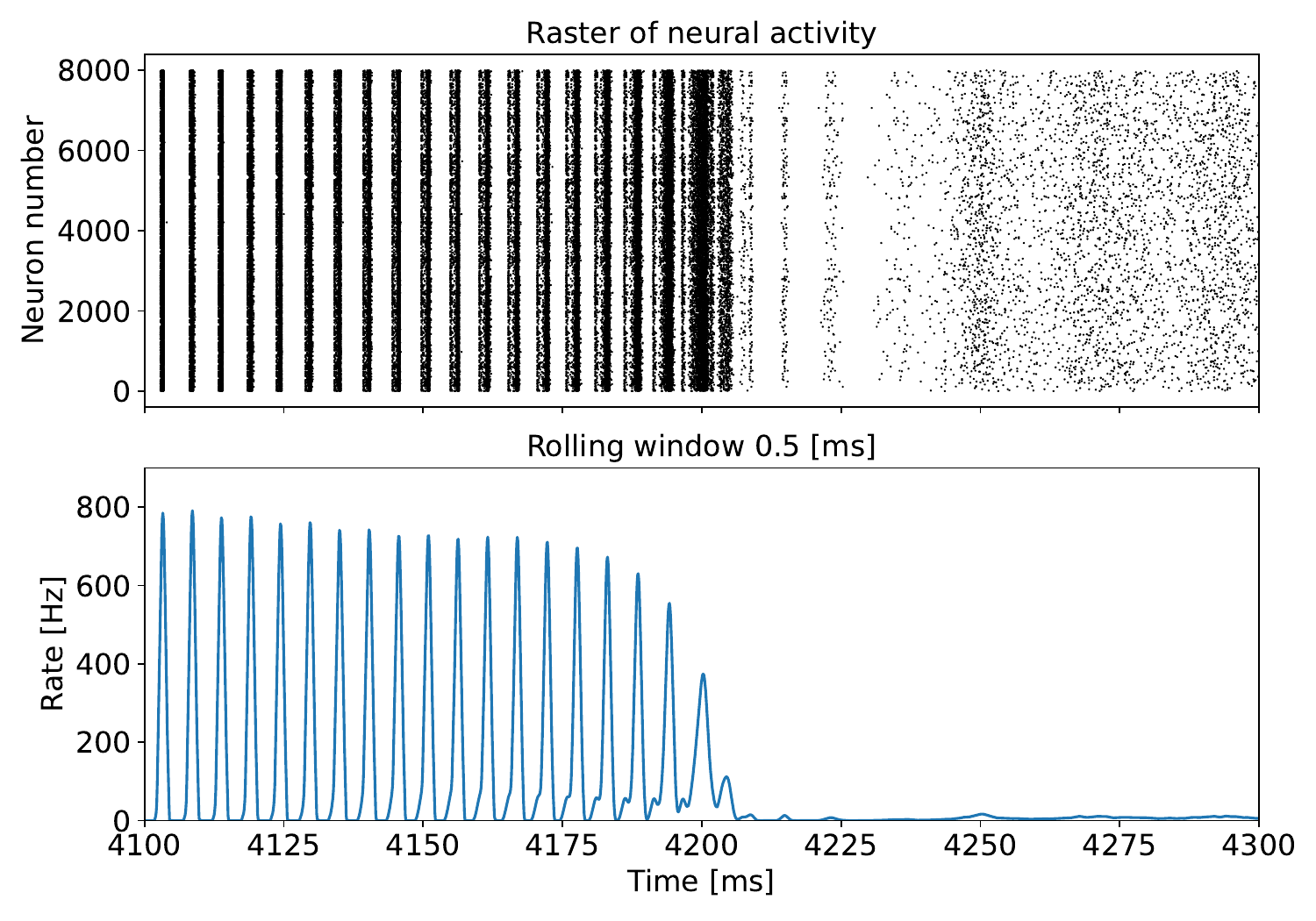}
	\textit{(c)}\includegraphics[width=0.95\textwidth]{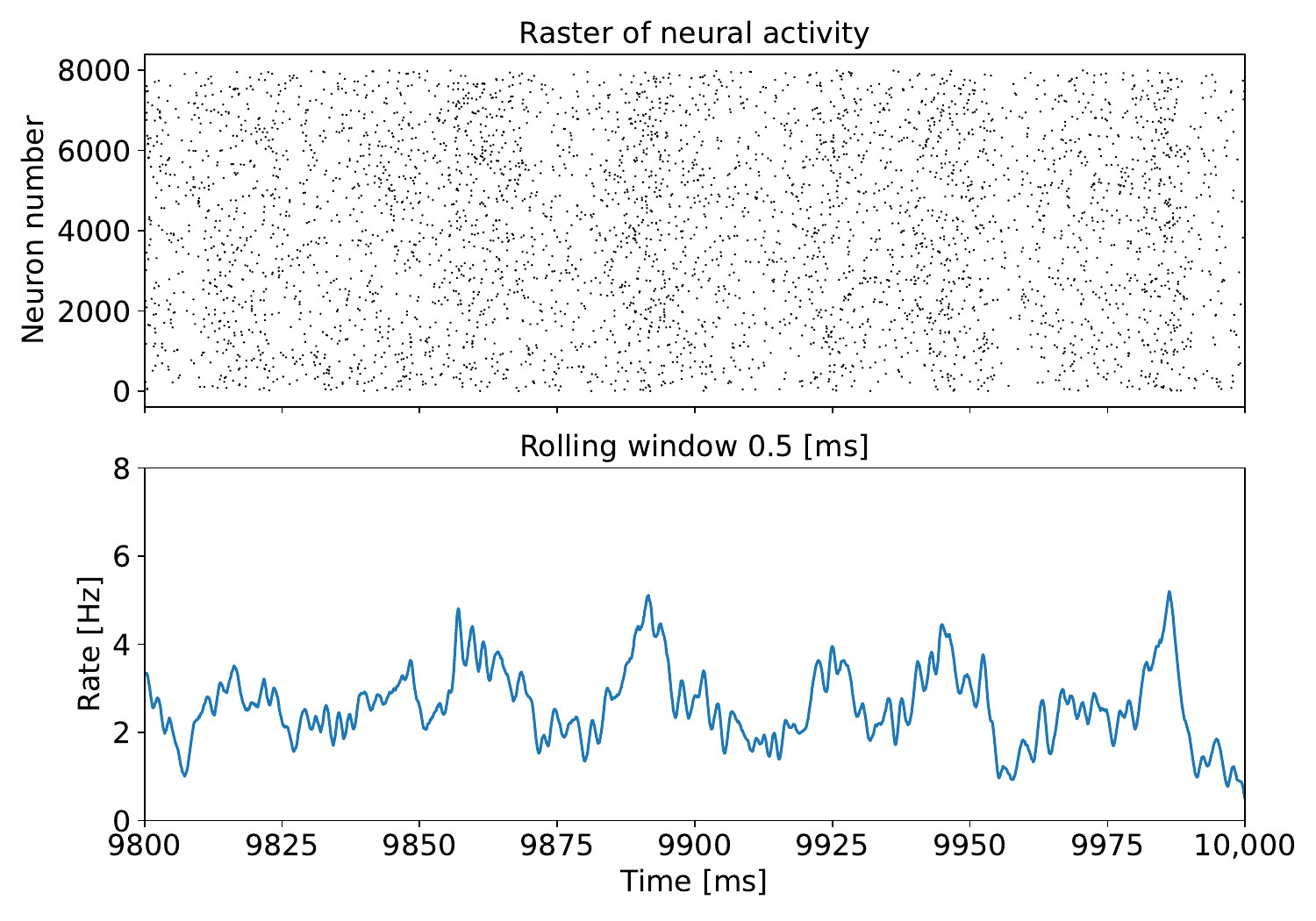}\\
	\caption{a) Diagram illustrating the synaptic contacts of spiking neuron network in the case with memristive plasticity. Excitatory neurons are highlighted in red, while inhibitory neurons are highlighted in blue. $w_{i,j}$- synaptic weight. Raster diagram illustrating neuronal activity and the population activity rate in the scenario involving memristor-based plasticity: (b) transition from synchronous to asynchronous modes, c) asynchronous modes at the end of simulation.}
	\label{fig:raster_stdp_only}
\end{figure}

Considering synaptic plasticity in the connections between inhibitory and excitatory neurons in the model of a spiking neuron network enables the brain to effectively regulate the dynamics of the neural network during the processing of a sensory signal. This contributes to a shift in the balance of excitation and inhibition, leading to the suppression of burst dynamics \cite{vogels2011inhibitory}. To incorporate such regulation into neuromorphic computing systems, a memristor can be employed to simulate synaptic plasticity \cite{stasenko2023control}. Consequently, we introduced memristor-based plasticity into the model in synaptic contacts between inhibitory and excitatory neurons (see Figure \ref{fig:raster_stdp_only}a). From the Figure \ref{fig:raster_stdp_only}b, it can be observed that during the interval from 4 to 5 seconds of the model simulation, in the presence of memristor-based plasticity, neuronal synchronization breaks down, and the burst mode disappears. The spiking neural network transitions into an asynchronous mode (as shown in the Figure \ref{fig:raster_stdp_only}c). 

\begin{figure}[H]
	\centering
	\includegraphics[width=0.95\textwidth]{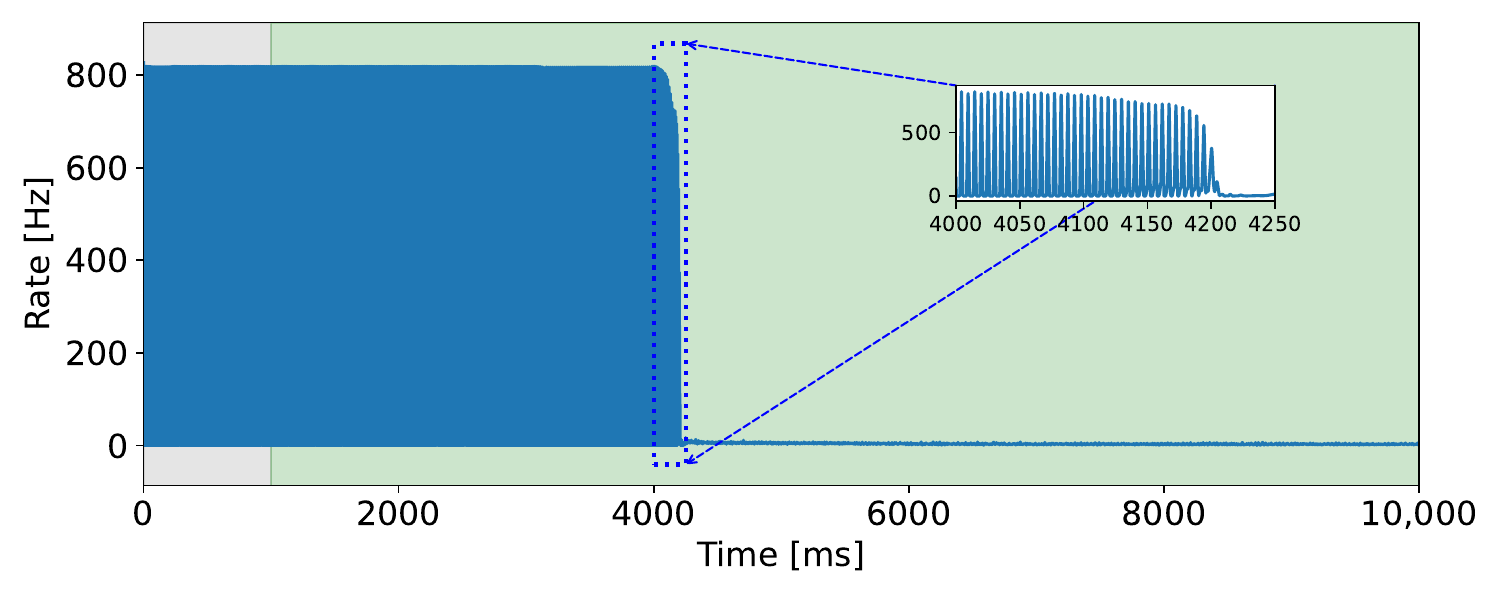}
	\caption{Population activity rate over a 10-second interval. The gray region signifies the absence of memristor-based plasticity, whereas the green region indicates its presence, respectively.}
	\label{fig:lfp_stdp_only}
\end{figure}

Notably, the disappearance of burst dynamics doesn't occur immediately but unfolds gradually over the course of memristor-based plasticity activity (as depicted in the Figure \ref{fig:lfp_stdp_only}). In the Figure \ref{fig:lfp_stdp_only}, the absence of memristor-based plasticity in the neural network is represented by the gray area in the first second, which is characterized by regular burst dynamics. Between 1 and 10 seconds, the model simulation occurs in the presence of memristor-based plasticity (depicted in the green area). As you can see, a regular burst mode in the neural network is observed from 1 to 4 seconds and undergoes a qualitative change during the interval from 4 to 5 seconds, with a reduction in the amplitude of burst discharges until they completely disappear, leading to asynchronous network activity.

\begin{figure}[H]
	\centering
	\textit{(a)}\includegraphics[width=0.95\textwidth]{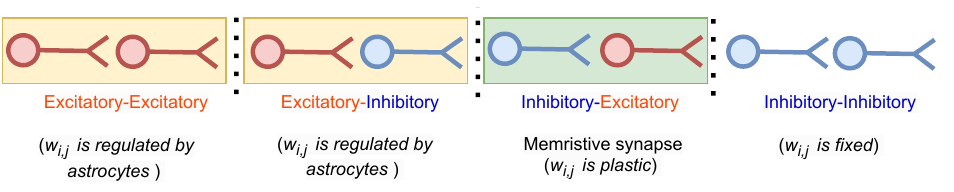}
	\textit{(b)}\includegraphics[width=0.95\textwidth]{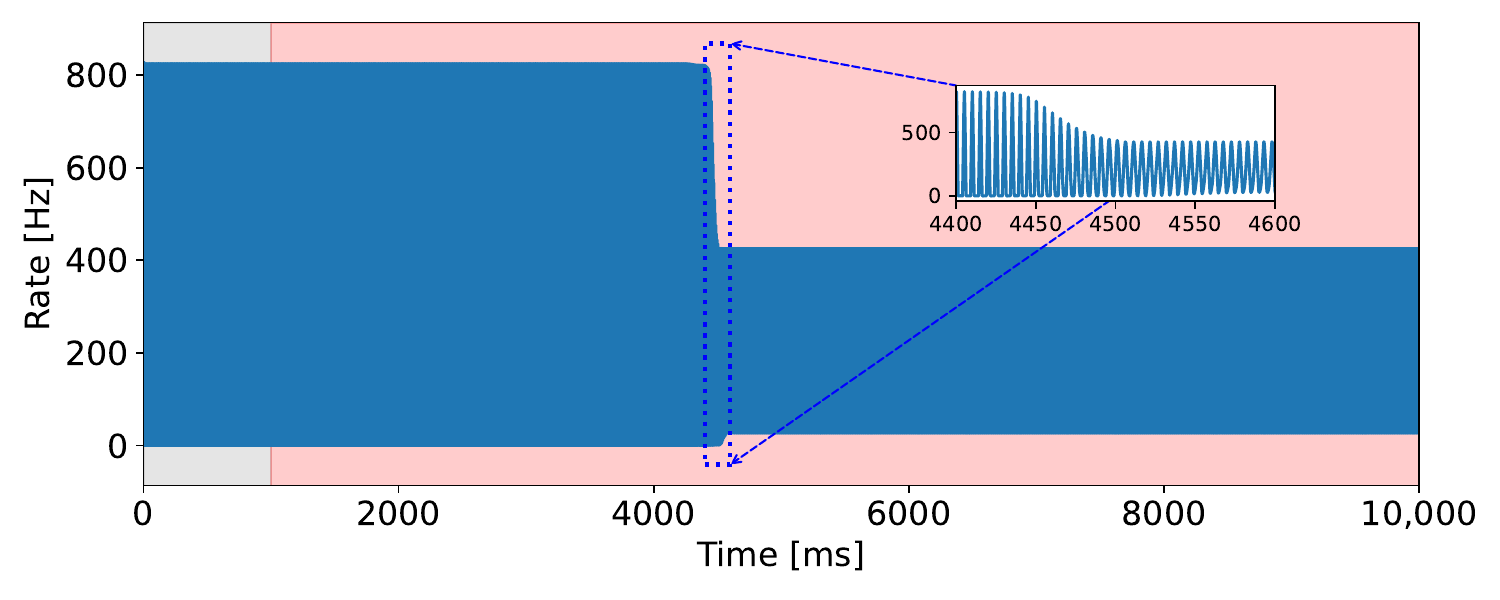}
	\caption{ a) Diagram illustrating the synaptic contacts of spiking neuron network in the case with memristor-based plasticity and astrocyte regulation. Excitatory neurons are highlighted in red, while inhibitory neurons are highlighted in blue. $w_{i,j}$- synaptic weight. b) Population activity rate over a 10-second period with astrocyte regulation of synaptic transmission. The gray region signifies the absence of memristor-based plasticity, while the red region indicates its presence, respectively.}
	\label{fig:lfp_stdp_astrocyte}
\end{figure}

Incorporating astrocytic regulation of synaptic transmission at glutamatergic (excitatory) synapses alters the dynamics of the neural network. The connection diagram of such a neural network is depicted in Figure \ref{fig:lfp_stdp_astrocyte}a. In our model, we specifically consider astrocytic regulation of excitatory (glutamatergic) synapses, as these have been the most extensively studied experimentally. As depicted in the Figure \ref{fig:lfp_stdp_astrocyte}b, the population activity rate undergoes changes in the presence of both astrocytic regulation and memristor-based plasticity. In the previous scenario, the gray area represents network activity in the absence of memristor-based plasticity, and the red area signifies its presence. In this case, astrocytic regulation is active throughout the entire simulation period. As shown in the Figure \ref{fig:lfp_stdp_astrocyte}b, similar to the second case, there is a long-term adaptation of the network and a significant change in the amplitude of burst discharges after 3 seconds of model simulation in the presence of memristor-based plasticity. Furthermore, burst dynamics persist in the presence of astrocytic regulation throughout the simulation.

\begin{figure}[H]
	\centering
	\includegraphics[width=0.95\textwidth]{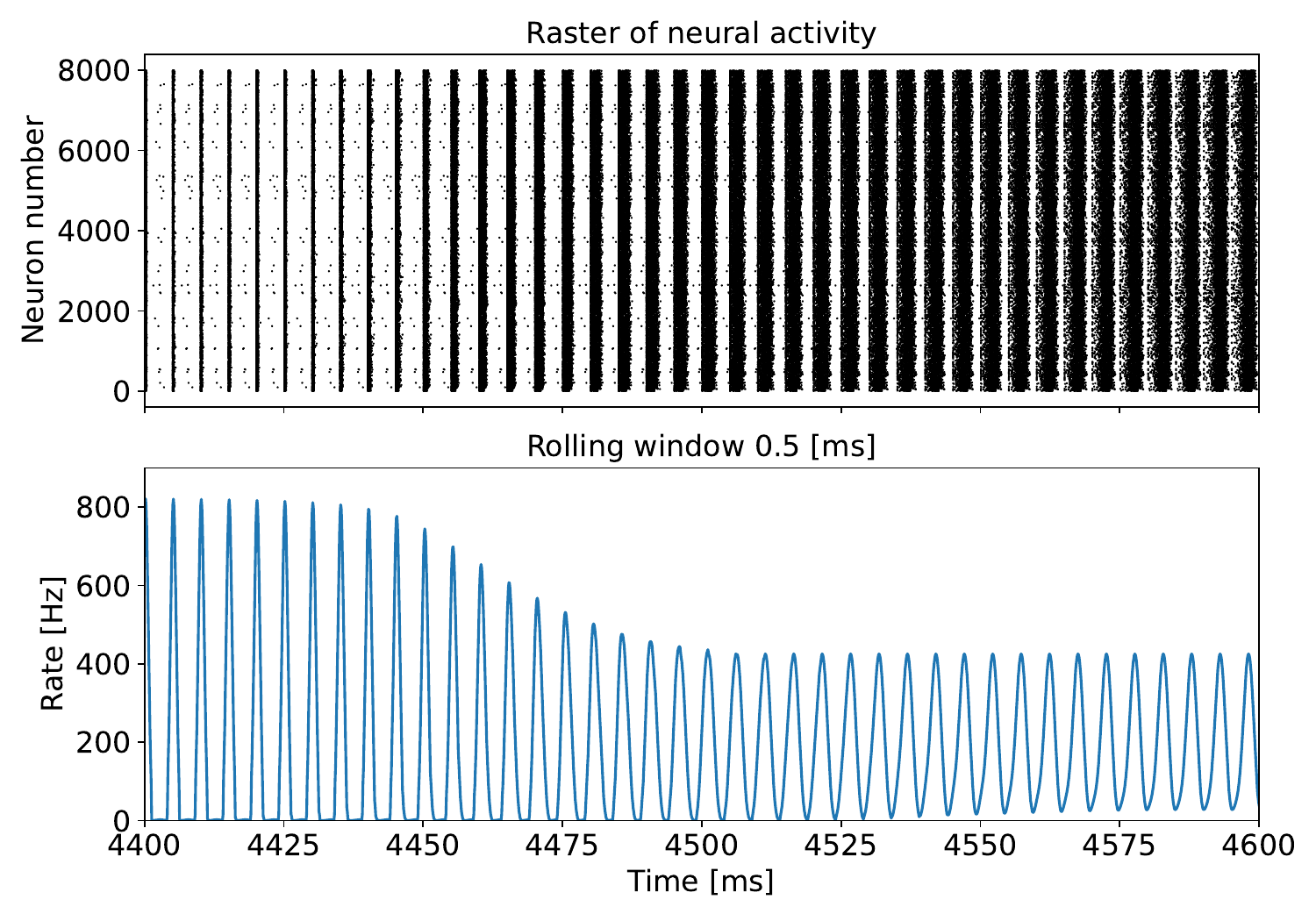}
	\caption{ Raster diagram illustrating neuronal activity and the population activity rate for the scenario involving memristor-based plasticity and astrocyte regulation of synaptic transmission during the period from 4.4 to 4.6 seconds.}
	\label{fig:raster_stdp_asto_transition}
\end{figure}

While observing the change in the amplitude of burst discharges from 4.4 to 4.6 seconds (as shown in the Figure \ref{fig:raster_stdp_asto_transition}), one can notice the formation of a small "tail" in the shape of burst stratification. This indicates a decrease in the amplitude of the population activity rate while, simultaneously, increasing their duration.

\begin{figure}[H]
	\centering
	\textit{(a)}\includegraphics[width=0.95\textwidth]{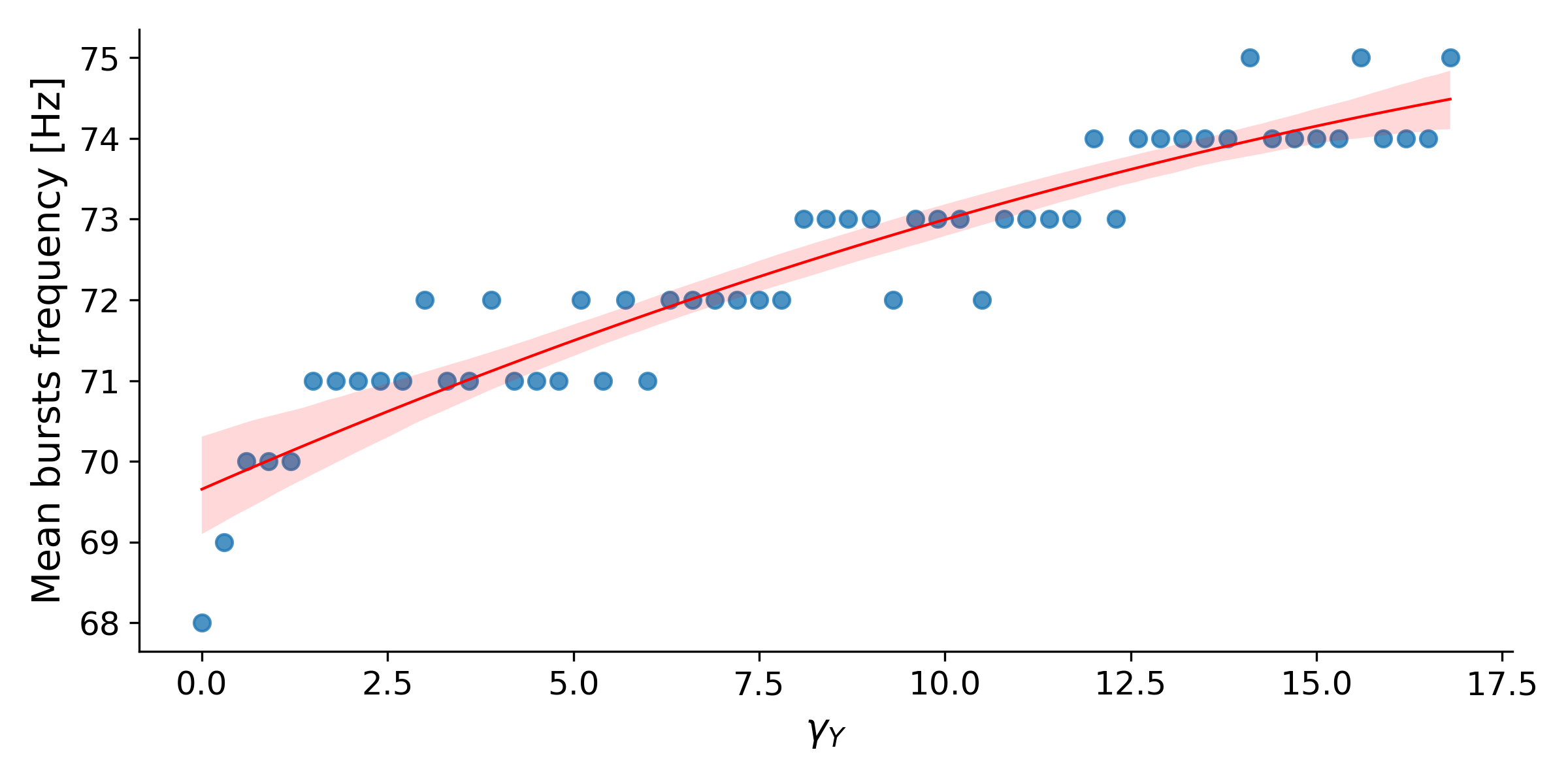}
	\textit{(b)}\includegraphics[width=0.95\textwidth]{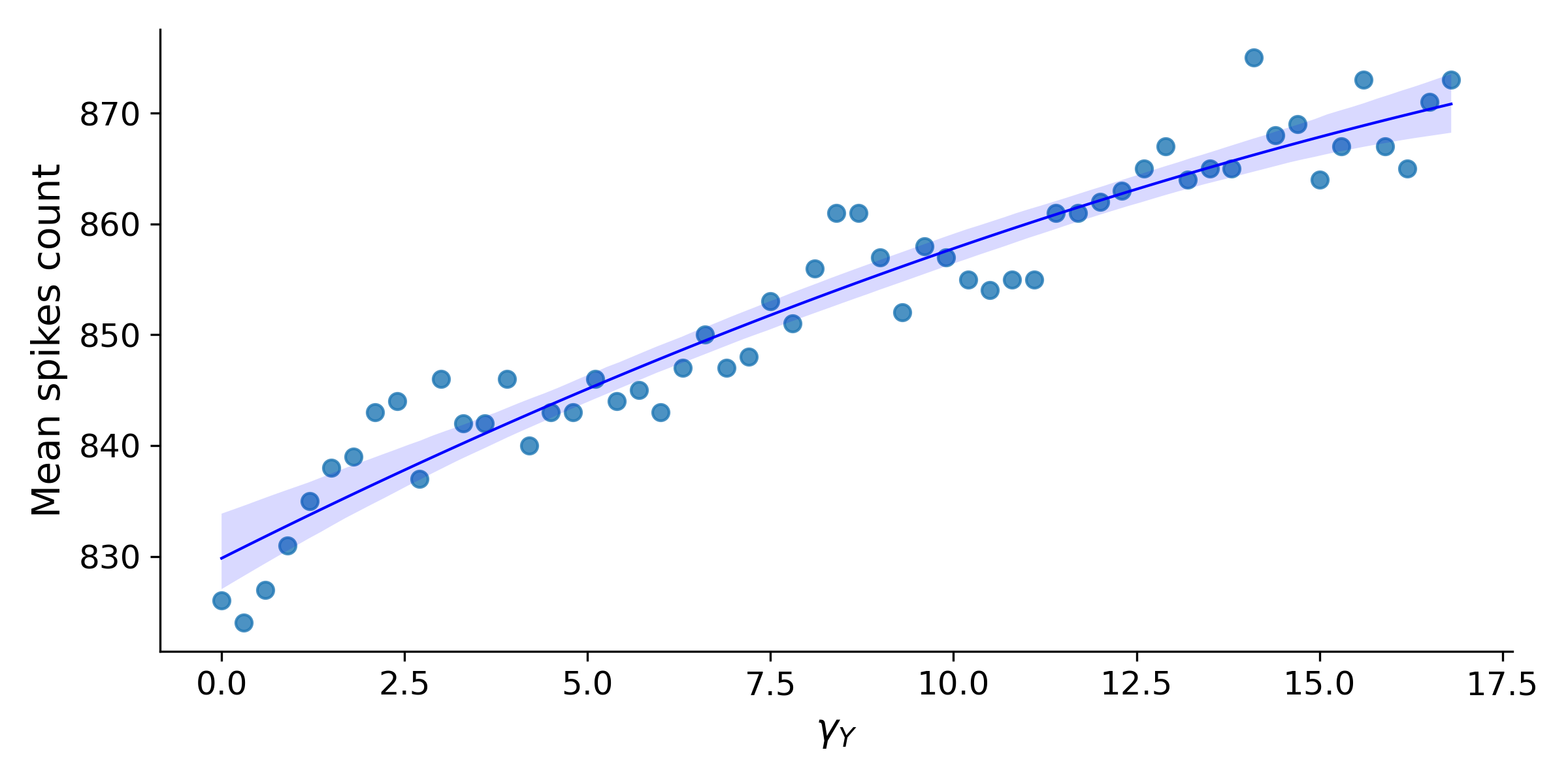}
	\caption{The association between the mean burst frequency (\textbf{a}) and the average number of spikes (\textbf{b}) per realization, as well as the extent of astrocytic influence on synaptic transmission (astrocytic potentiation) denoted by $\gamma_Y$, within the context of memristor-based plasticity.}
	\label{fig:influence_astrocyte}
\end{figure}

We next examined the effect of astrocytic plasticity on the mean frequency of burst generation (Figure \ref{fig:influence_astrocyte}a) and the mean number of spikes (Figure \ref{fig:influence_astrocyte}b) in the presence of memristor-based plasticity. To do this, we considered temporary implementations lasting 10 seconds. For 5 experiments, the values were then averaged and entered into the figure. When calculating the average number of spikes and the average frequency of burst generation, we did not take into account the first second of the model simulation, since transient processes could occur during this period.

You can notice that as the parameter, $\gamma_Y$, increases, both the burst frequency and the number of spikes increase. The non-zero first value is explained by the fact that there is a transition period of the influence of memristive plasticity, which takes some time. 

\begin{figure}[H]
	\centering
	\textit{(a)}\includegraphics[width=0.95\textwidth]{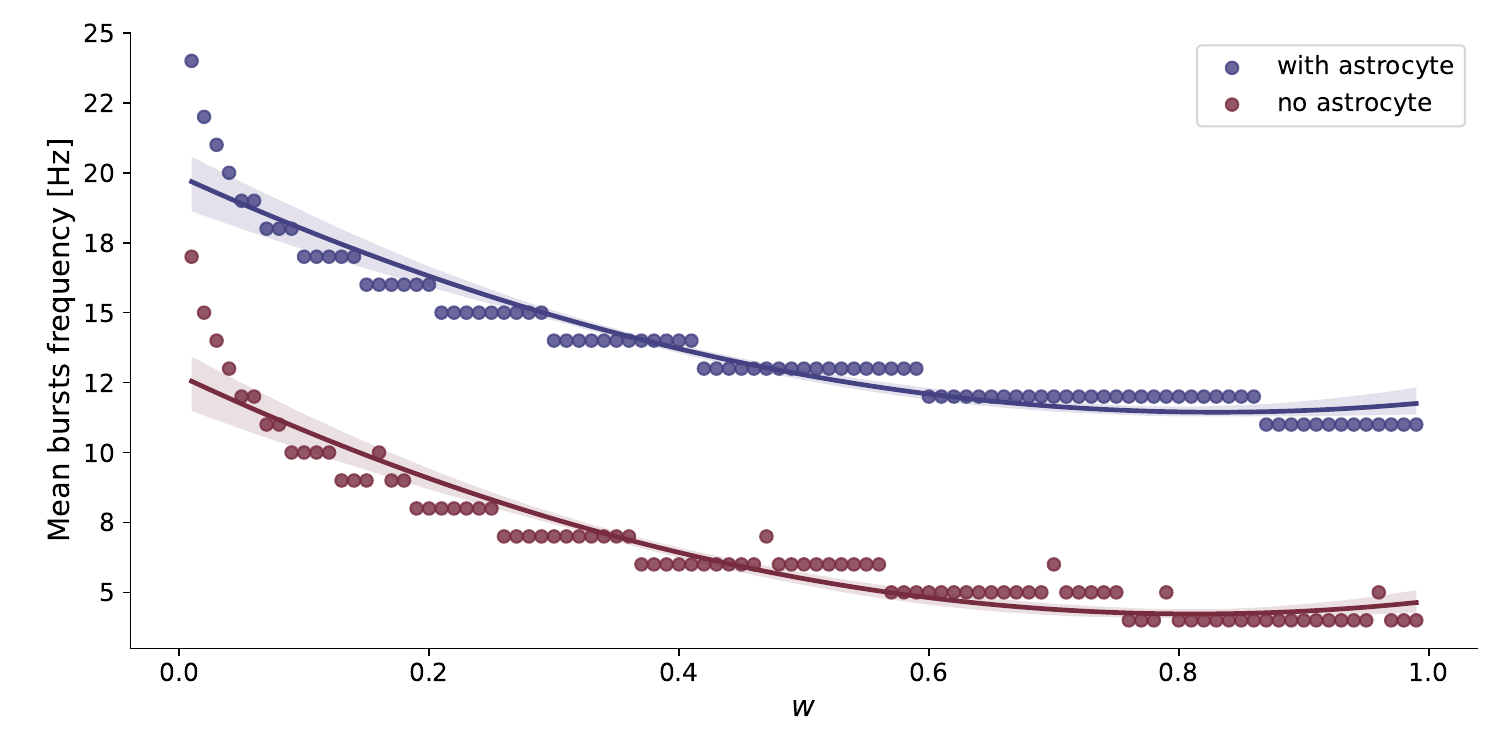}
	\textit{(b)}\includegraphics[width=0.95\textwidth]{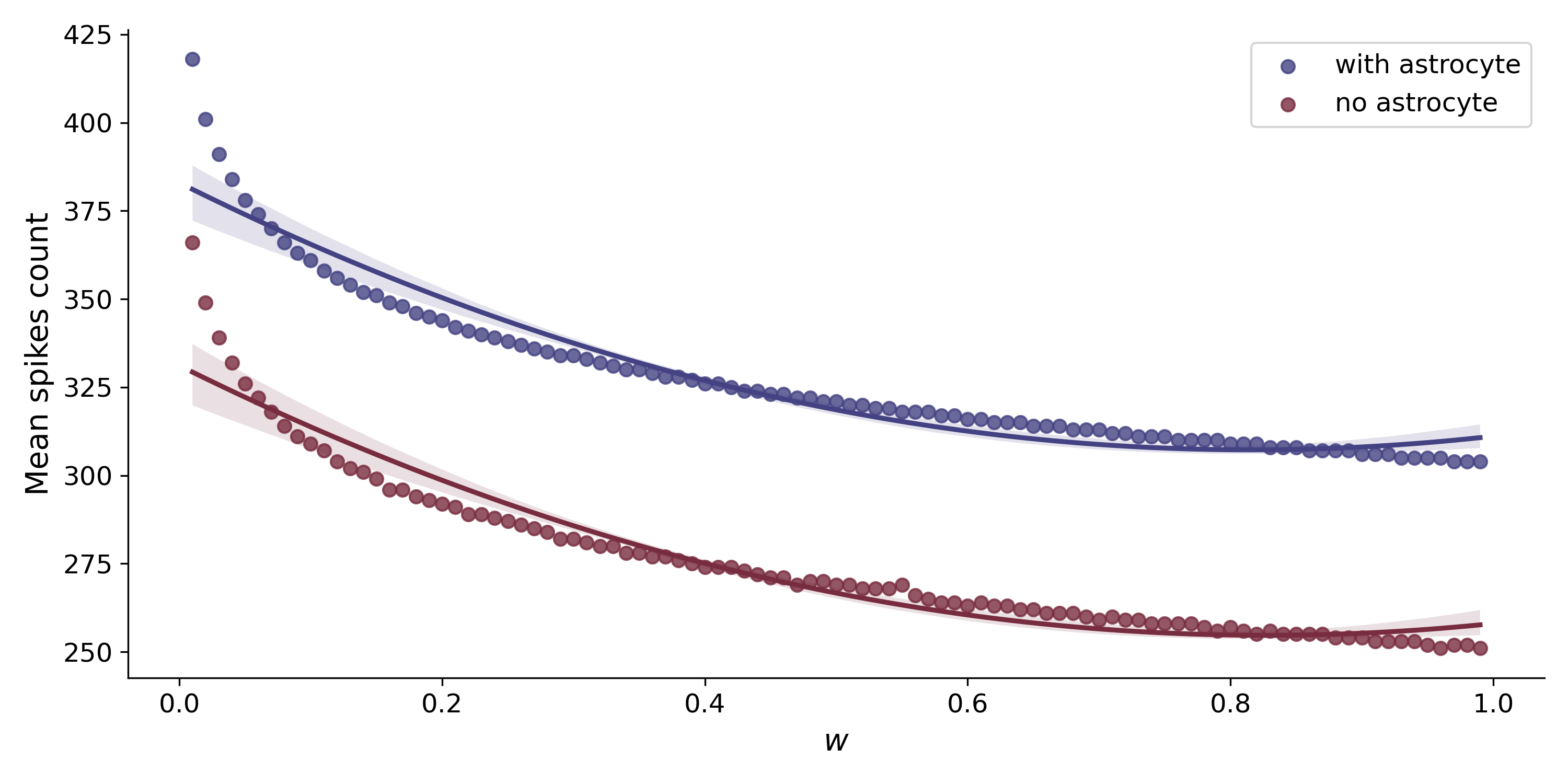}
	\caption{Relationship between the average burst frequency (\textbf{a}) and the average number of spikes (\textbf{b}) for different values of the synaptic weight $w$ of a memristive synapse in the absence (red dots and red regression curve) and presence (blue dots and blue regression curve) of astrocytic modulation of synaptic transmission.}
	\label{fig:influence_w}
\end{figure}

The influence of the synaptic weight $w$ of the memristive synapse in the absence and presence of astrocytic modulation of synaptic transmission on the average burst frequency and number of spikes was also investigated (Figure \ref{fig:influence_w}).

As can be observed, astrocytic modulation of synaptic transmission significantly increases both burst frequency and the total number of spikes at different values of the synaptic weights of the memristive synapse.

\section{Discussion}
\label{sec:discussion}

The implications drawn from the examination of bursting dynamics in spiking neural networks (SNNs) go beyond the realm of neuroscience. Computational models based on bursting are highly valuable for neuromorphic computing, providing energy-efficient and biologically plausible paradigms. Recent research endeavors involve the comparison of spiking and bursting dynamics in liquid computing and find that bursting networks exhibit superior computational performance, expanded information capacity, improved synaptic integration efficiency, enhanced activity complexity, and proficiency in information processing \cite{li2016bursting}. These findings underscore the efficiency of bursting for signal processing and computations.

The interaction among synaptic plasticity mechanisms, inhibitory circuits, astrocyte regulation of synaptic transmission, and memristor-based plasticity \cite{stasenko2023control} presents a comprehensive approach to modulating network bursting in spiking neural networks (SNNs). This approach aims to fully harness their capabilities for efficient information processing and cognitive applications. Memristive devices show promise in achieving efficient and scalable computing systems that emulate neural network functionalities \cite{yang2013memristive}. Their ability to facilitate adaptive synaptic weight adjustments makes them valuable for applications in neuromorphic computing and various domains \cite{zafar2022classifying, wang2021artificial, stasenko2023model}.

To date, work has been carried out on the development of an analog circuit for the Li-Rinzel calcium model based on 180 nm CMOS technology \cite{khosravi2022new}. The presented analog circuit demonstrates low power consumption. Utilizing analog chips and neuromorphic hardware models with ultra-low power consumption and fast processing, neuron-glial interactions \cite{barros2021engineering,bicaku2023power}, which are crucial in the regulation of neural activity and information transmission processes in biological systems, were also simulated.

Existing work demonstrates the feasibility of implementing LIF neurons and networks in hardware using CMOS technology \cite{asghar2021low}, memristor-like STDP models in nanocomposite memristors \cite{demin2021necessary}, along with astrocyte processes and neuron-glial interaction, suggesting the potential for realizing the proposed model in hardware soon.

One limitation of the proposed model can be considered the use of a simplified neuron model, and the modeling of astrocytic dynamics and neuron-glial interaction was achieved using a mean-field approach. This approach allows for a phenomenological description of the observed experimental effects without the need for detailed biophysical modeling. Such an approach significantly improves the computational efficiency of the model without sacrificing biological relevance, which is crucial for the development of neuromorphic computing systems and devices.

The enhancements in the model are a result of incorporating memristive regulation, which influences the dynamics of spiking neuron networks in the execution of brain information functions.


\section{Conclusion}
\label{sec:conclusion}
In conclusion, we have proposed a new model for controlling the burst dynamics of a spiking neural network with memristor-mediated plasticity through astrocytic modulation of synaptic transmission at excitatory (glutamatergic) synapses. In the absence of memristor-mediated plasticity, connections between these neurons drive network dynamics into an explosive mode, as observed in many experiments involving dissociated neuronal cultures. However, incorporating memristive plasticity, which mimics synaptic plasticity at inhibitory synapses, into inhibitory connections results in a shift in network dynamics toward a single asynchronous firing mode. Subsequent inclusion of glutamatergic synapses in the model of astrocytic regulation makes it possible to restore burst dynamics. We conducted a study to investigate the control parameters of the influence of astrocytic regulation on the formation of packet dynamics within the network. These results represent a significant contribution to the understanding of the mechanisms governing neural network dynamics, which has enormous implications for problems related to information encoding and the development of neuromorphic computing systems.

\vspace{6pt} 



\authorcontributions{Conceptualization, S.V.S.; methodology, S.V.S.; software, S.V.S.; validation, S.V.S.; formal analysis, S.V.S.; investigation, S.V.S.; resources, S.V.S.; data curation, S.V.S.; writing—original draft preparation, S.V.S., A.N.M., A.A.F, V.A.S. and V.B.K.; writing—review and editing, S.V.S., A.N.M., A.A.F, V.A.S. and V.B.K.; visualization, S.V.S.; supervision, S.V.S.; project administration, S.V.S. and V.B.K.; funding acquisition, S.V.S. and V.B.K. All authors have read and agreed to the published version of the manuscript.
.}

\funding{The work regarding the development of a mathematical model of a spiking neural-glial network was supported by the Russian Science Foundation (Project No. 23-11-00134); the work regarding the implementation of the memristive plasticity model and the selection of its parameters was supported by a grant from the Russian Federation Government (Agreement No. 075-15-2022-1123).}

\institutionalreview{Not applicable.}

\informedconsent{Not applicable.}

\dataavailability{The data that support the findings of this study are available from the corresponding author upon reasonable request.} 


\conflictsofinterest{The authors declare no conflict of interest.} 



\appendixtitles{yes} 
\appendixstart
\appendix

\begin{adjustwidth}{-\extralength}{0cm}

\reftitle{References}

\end{adjustwidth}
\end{document}